\documentclass[aps,prl,twocolumn,10pt,superscriptaddress,amsfonts,amssymb,amsmath,preprintnumbers,floatfix,showpacs]{revtex4-1}
\usepackage{graphicx}
\usepackage{bm}
\usepackage{color,ulem}

\begin{document}
\title{Observation of a dispersive charge mode in hole-doped cuprates using resonant inelastic x-ray scattering at the oxygen $K$ edge}
\author{K. Ishii}
\affiliation{Synchrotron Radiation Research Center,
National Institutes for Quantum and Radiological Science and Technology, Hyogo 679-5148, Japan}
\author{T. Tohyama}
\affiliation{Department of Applied Physics, Tokyo University of Science, Tokyo 125-8585, Japan}
\author{S. Asano}
\affiliation{Institute for Materials Research, Tohoku University, Sendai 980-8577, Japan}
\author{K. Sato}
\affiliation{Institute for Materials Research, Tohoku University, Sendai 980-8577, Japan}
\author{M. Fujita}
\affiliation{Institute for Materials Research, Tohoku University, Sendai 980-8577, Japan}
\author{S. Wakimoto}
\affiliation{Materials Science Research Center, Japan Atomic Energy Agency, Ibaraki 319-1195, Japan}
\author{K. Tustsui}
\affiliation{Synchrotron Radiation Research Center,
National Institutes for Quantum and Radiological Science and Technology, Hyogo 679-5148, Japan}
\author{S. Sota}
\affiliation{Computational Materials Science Research Team,
RIKEN Advanced Institute for Computational Science (AICS), Kobe, Hyogo 650-0047, Japan}
\author{J. Miyawaki}
\affiliation{Institute for Solid State Physics, University of Tokyo, Chiba 277-8581, Japan}
\author{H. Niwa}
\affiliation{Institute for Solid State Physics, University of Tokyo, Chiba 277-8581, Japan}
\author{Y. Harada}
\affiliation{Institute for Solid State Physics, University of Tokyo, Chiba 277-8581, Japan}
\author{J. Pelliciari}
\altaffiliation{Present address: Department of Physics, Massachusetts Institute of Technology, Cambridge, MA 02139, USA}
\affiliation{Research Department Synchrotron Radiation and Nanotechnology, Paul Scherrer Institut, CH-5232 Villigen PSI, Switzerland}
\author{Y. Huang}
\affiliation{Research Department Synchrotron Radiation and Nanotechnology, Paul Scherrer Institut, CH-5232 Villigen PSI, Switzerland}
\author{T. Schmitt}
\affiliation{Research Department Synchrotron Radiation and Nanotechnology, Paul Scherrer Institut, CH-5232 Villigen PSI, Switzerland}
\author{Y. Yamamoto}
\affiliation{Graduate School of Science and Technology, Kwansei Gakuin University, Hyogo 669-1337, Japan}
\author{J. Mizuki}
\affiliation{Graduate School of Science and Technology, Kwansei Gakuin University, Hyogo 669-1337, Japan}
\date{\today}

\begin{abstract}
We investigate electronic excitations in La$_{2-x}$(Br,Sr)$_x$CuO$_4$ using resonant inelastic x-ray scattering (RIXS) at the oxygen $K$ edge.
RIXS spectra of the hole-doped cuprates show clear momentum dependence below 1 eV.
The spectral weight exhibits positive dispersion and shifts to higher energy with increasing hole concentration.
Theoretical calculation of the dynamical charge structure factor on oxygen orbitals in a three-band Hubbard model is consistent with the experimental observation of the momentum and doping dependence, and therefore the dispersive mode is ascribed to intraband charge excitations which have been observed in electron-doped cuprates.
\end{abstract}

\pacs{71.27.+a,74.25.Jb,74.72.Gh,78.70.Ck}
\preprint{preprint \today}

\maketitle

Strongly correlated transition-metal oxides display various interesting physical properties including metal-insulator transition, high-temperature superconductivity and colossal magnetoresistance, and some of the oxides are classified into doped Mott insulators where electron correlation significantly modifies their band structure which diverges from that of the noninteracting simple metal \cite{Imada1998}.
Among the doped Mott insulators, superconducting cuprates are most intensively studied \cite{Lee2006}.
This is mainly due to the superconductivity at high transition temperature and related phenomena such as pseudogap and a competing phase with charge order \cite{Keimer2015}.
In addition to the interest of superconductivity, doped cuprates are important and suitable for the study of electronic structure of the doped Mott insulator because relatively simple theoretical models with a few orbitals are applicable to describe the electronic structure near the Fermi energy.
They represent the benchmark of the doped Mott insulators and the clarification of the fundamental electronic structure is essential for understanding the mechanism of the physical phenomena in the doped cuprates.

\begin{figure*}[t]
\centering
\includegraphics[width=1.0\textwidth]{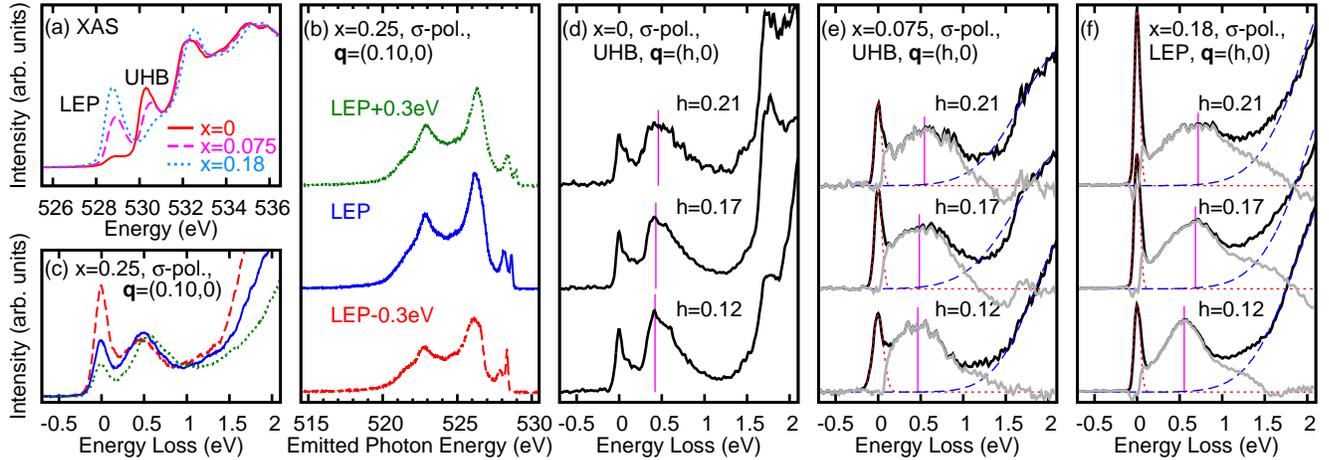}
\caption{(color online). (a) X-ray absorption spectra near the oxygen $K$-edge.
X-ray polarization is parallel to the CuO$_2$ plane.
(b) O $K$-edge RIXS spectra of La$_{2-x}$Sr$_x$CuO$_4$ ($x$ = 0.25) plotted against emitted photon energy.
From bottom to top, incident photon energy is tuned to LEP$-$0.3 eV, LEP, and LEP$+$0.3 eV of XAS.
(c) The same spectra as in (b) plotted as a function of energy loss.
(d-f) RIXS spectra of La$_{2-x}$Sr$_x$CuO$_4$.
The energy of the $\sigma$-polarized incident x-rays is tuned to UHB for $x$ = 0 (d) and $x$ = 0.075 (e) and LEP for x = 0.18 (f).
Black lines are the raw spectra and gray lines in (e) and (f) are the spectral weight after subtraction of the elastic peak (dotted
line) and high-energy tail (dashed line).
The vertical bars indicate the peak position of the spectral weight after subtraction.
}
\label{fig1}
\end{figure*}

In the undoped cuprates, only the spin degree of freedom contributes to the low-energy electron dynamics.
When carriers are doped, the charge degree of freedom becomes active and the electron dynamics is characterized by the motion of spin and charge.
Therefore, we consider that both spin and charge excitations must be investigated on equal footing in order to understand the electron dynamics characterizing the physics of the cuprates.
Inelastic neutron scattering (INS) has been widely used for studying the spin dynamics in the reciprocal lattice space, and high-resolution resonant inelastic x-ray scattering (RIXS) at the Cu $L_3$-edge has recently become an alternative to measure momentum-resolved spin excitations up to several hundreds meV \cite{Ament2009b,Braicovich2010}.
Charge excitations in the doped cuprates extend to higher energy than the spin excitations and the electron correlation affects the charge excitations of the order of a few eV.
Optical studies \cite{Uchida1991,Onose2004} demonstrated that spectral weight of the intraband charge excitations emerges below the charge-transfer gap in the doped cuprates and the weight characterizes the charge excitations in the doped Mott insulators.
The high-energy part of momentum-resolved charge excitations below the gap was studied by Cu $K$-edge RIXS \cite{Kim2004c,Ishii2005b,Wakimoto2013} but the huge tail of the elastic scattering hampers the observation below several hundreds meV.
Alternatively, the capability of Cu $L_3$-edge RIXS for the detection of the charge excitations is argued theoretically \cite{Jia2016,Tsutsui2016}, and particle-hole charge excitations have been reported in a Ti $L_3$-edge RIXS study on the weakly-correlated broadband material $1T$-TiSe$_2$ \cite{Monney2012}.
Nevertheless, momentum-dependent charge excitations in the hole-doped cuprates have, to the best of our knowledge, not been identified experimentally using the Cu $L_3$-edge RIXS.
On the other hand, in the electron-doped cuprates, a dispersive mode which is located at higher energy than the spin excitations was found in the Cu $L_3$-edge RIXS spectra \cite{Ishii2014,Lee2014} and it is ascribed to the particle-hole charge excitations \cite{Ishii2014}.
However, the origin of this mode is still controversial.
In Ref.~\cite{Lee2014}, a different interpretation that is associated with a symmetry-breaking state other than superconductivity
is proposed and this dispersive mode is considered to be absent in hole-doped cuprates. 
Furthermore, it is discussed in a theoretical study \cite{Greco2016} that the dispersive mode comes from a plasmon excitation.
In order to settle this controversy, it is necessary to confirm whether such a dispersive mode exists also in the hole-doped cuprates.

In this Letter, we report that a dispersive charge mode in the hole-doped cuprates can be observed using O $K$-edge RIXS.
Because doped holes predominantly occupy the O $2p$ orbitals in the cuprates, the O $K$-edge RIXS represents a direct way to probe the charge dynamics of doped carriers and the intensity of charge modes can be expected to be sizable.
The experimental observation of the momentum and doping dependence is consistent with a theoretical calculation of the dynamical charge structure factor on oxygen orbitals in a three-band Hubbard model with Cu3$d_{x^2-y^2}$ and two O2$p_\sigma$ orbitals.
Therefore, we ascribe the dispersive mode to intraband charge excitations.
Our result demonstrates that the intraband charge excitations are qualitatively symmetric between the hole- and electron-doped cuprates.

\begin{figure*}[t]
\centering
\includegraphics[width=1.0\textwidth]{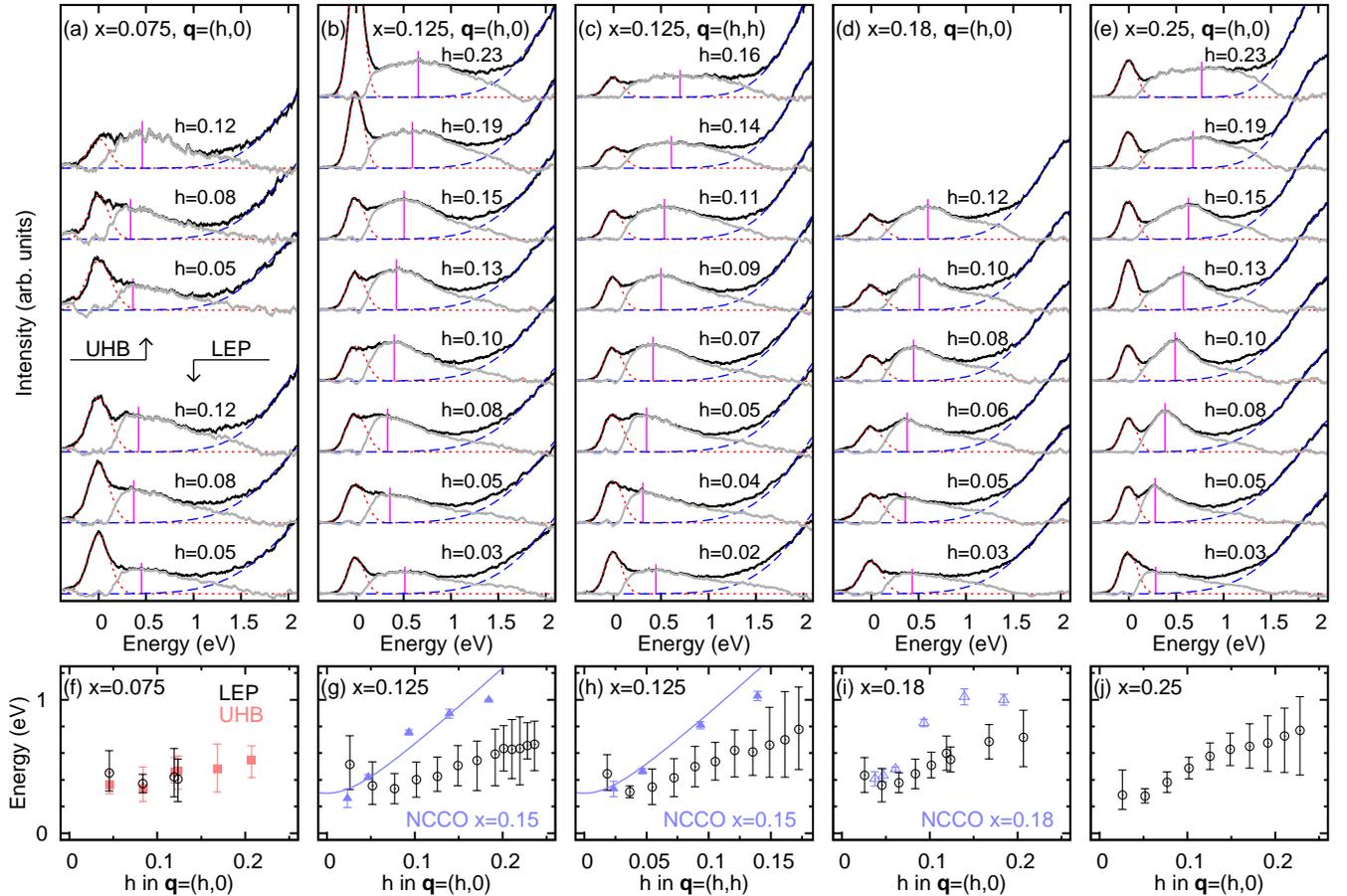}
\caption{(color online). (a-e) RIXS spectra of La$_{2-x}$(Br,Sr)$_x$CuO$_4$.
The Energy of the $\sigma$-polarized incident x-rays is tuned to LEP of XAS except for the upper three spectra in (a).
Black lines are the raw spectra and gray lines are the spectral weight after subtraction of the elastic peak (dotted
line) and high-energy tail (dashed line).
The vertical bars indicate the peak position of the spectral weight after subtraction.
(f-j) Dispersion relation of the dispersive mode.
Peak positions are plotted by open circles and filled squares.
Full widths at the half maximum of the peaks are displayed by vertical bars.
Results from the spectra in Figs.~\ref{fig1} (c) and (d) are included.
Peak positions of the charge excitations \cite{Ishii2014} and the fast-dispersive mode \cite{Lee2014} in electron-doped Nd$_{2-x}$Ce$_x$CuO$_4$ are also shown by triangles and solid lines, respectively.
%The filled circle in (j) indicates the peak of intraband charge excitations in LSCO ($x$ = 0.30) observed in the Cu $K$-edge RIXS \cite{Wakimoto2013}.
}
\label{fig2}
\end{figure*}

The RIXS experiments were performed using the SAXES spectrometer \cite{Ghiringhelli2006} at the ADRESS beam line \cite{Strocov2010} of the Swiss Light Source (SLS)  at the Paul Scherrer Institut and the HORNET spectrometer \cite{Harada2012} at beam line BL07LSU \cite{Senba2011} of SPring-8.
The RIXS spectra in Figs.~\ref{fig1}(d)-(f) were taken at the former and others were measured at the latter, with respective energy resolutions of 60 and 170-190 meV.
The scattering angle ($2\theta$) was set to 130$^{\circ}$ for Figs.~\ref{fig1}(d,e,f), 90$^{\circ}$ for Figs.~\ref{fig2}(a,d), and 135$^{\circ}$ for Figs.~\ref{fig1}(b,c)  and Figs.~\ref{fig2}(b,c,e).
Single crystals of La$_{2-x}$Sr$_x$CuO$_4$ (LSCO with $x$ = 0, 0.075, 0.18, and 0.25) and La$_{2-x}$Ba$_x$CuO$_4$ (LBCO with $x$ = 0.125) were measured at the base temperature of spectrometers (10-30 K).
Because the difference of alkaline earth metals (Sr and Ba) is not important here, we distinguish the samples by hole concentration ($x$). 
The crystals were cleaved before the measurement and $\sigma$-polarized x-rays were irradiated on the $ab$-plane of the crystals.
The $c$-axis was kept parallel to the horizontal scattering plane and momentum transfer in the CuO$_2$ plane ($\mathbf{q}$) was scanned by rotating the crystal along the vertical axis.

In the x-ray absorption spectrum (XAS) of the doped cuprates, two peaks are observed near the O $K$ edge \cite{Chen1991}.
As shown in Fig.~\ref{fig1}(a), the spectral weight of the transition to the upper Hubbard band (UHB) transfers to the lower-energy peak (LEP) with increasing the hole concentration.
The incident photon energy for the RIXS measurements was tuned to the top of either LEP or UHB of respective samples.
In the undoped $x$ = 0, large enhancement of excitations is observed only at the UHB resonance.
When holes are doped to $x$ = 0.075, the spectral weight of XAS becomes comparable between LEP and UHB.
We measured the RIXS spectra at both LEP and UHB resonances, but, as shown in Fig.~\ref{fig2}(a,f), we could not find any significant difference at the sub-eV range between the two resonance conditions.
Because the spectral weight of LEP is dominant in XAS for higher doping ($x$ = 0.125, 0.18, and 0.25), we took the RIXS spectra at the LEP resonance .
Figure \ref{fig1}(b) shows the incident photon energy ($E_i$) dependence near LEP for $x$ = 0.25.
The spectral shape is almost unchanged below 527 eV, meaning that most of the spectral weight in this energy region comes from fluorescence.
The spectra discussed below are normalized to the integrated intensity of the fluorescence.
In Fig.~\ref{fig1}(c), we plot the same spectra as in Fig.~\ref{fig1}(b) as a function of energy loss.
The peak around 0.5 eV slightly shifts to higher energy with increasing $E_i$, indicating a non-perfect Raman behavior, but the shift is much smaller than the variation of $E_i$.

Figures \ref{fig1}(d)-(f) show the momentum dependence of the O $K$-edge RIXS spectra for $x$ = 0, 0.075, and 0.18, respectively.
A clear peak is observed below 1 eV for all three samples.
The peak position and the lineshape of the undoped compound ($x$ = 0) are independent of $\mathbf{q}$ and the spectral weight of the peak is ascribed to two-magnon excitations \cite{Harada2002,Bisogni2012a}.
In contrast, the peak position of $x$ = 0.075 and 0.18 changes with momentum.
The qualitative difference of the momentum dependence indicates that another type of excitations exists in the spectra of the doped compounds.

For the doped compounds, we measured the RIXS spectra in a finer interval of momentum and summarize them in Figs.~\ref{fig2}(a)-(e).
We subtract the elastic scattering and high-energy tail from the raw spectra and plot the peak positions and widths of residual intensity (gray lines) as a function of $\mathbf{q}$ in Figs.~\ref{fig2}(f)-(j).
Except for the close vicinity of $\mathbf{q}=(0,0)$, the spectral weight at the sub-eV region shifts to higher energy with increasing momentum transfer, forming a dispersive mode.
The magnitude of the dispersion of the mode becomes larger with increasing hole concentration.
For example, the peak position at $\mathbf{q}=(0.23,0)$ is 0.66 eV for $x=0.125$ while it is 0.77 eV for $x=0.25$.
Comparing the spectral weights at the same $\vert \mathbf{q} \vert$ of $x=0.125$, the peak of the $(h,h)$ direction is located at higher energy than that of the $(h,0)$ direction.

In a published work on LSCO, the peak at the sub-eV region in the O $K$-edge RIXS spectra was ascribed to two-magnon excitations not only for the undoped compound but also for the doped ones \cite{Bisogni2012a,Bisogni2012b}.
We do not exclude the possibility that a part of the spectral weight of the doped compounds comes from two-magnon excitations, but the dispersive mode should have another origin from the following reasons.
Firstly, the energy of the peak positions is too high to ascribe the dispersive mode to two-magnons.
The excitation energy at large $\mathbf{q}$ in the present measurements is 0.6-0.8 eV while calculated two-magnon density of states for a nearest neighbor Heisenberg antiferromagnet with exchange interaction $J$ is located at lower than $4J \simeq 0.5$ eV \cite{Forte2008}.
Secondly, in Raman scattering studies \cite{Sugai1988,Ruebhausen1997,Naeini1999,Li2013}, the energy of two-magnon excitations rapidly decreases upon hole doping in contrast with our experimental fact that the peak of the dispersive mode at large $\mathbf{q}$ shifts to higher energy with increasing hole concentration.
Thirdly, the peak becomes salient in the overdoped compound ($x=0.25$) which is situated farthest from the antiferromagnetic ordered phase.
Finally, theoretical calculations of two-magnon excitations in a hole-doped $t$-$J$-type model show negligible $\mathbf{q}$ dependence (see Supplemental Material S1~\cite{Supple}).
All of these facts suggest that two-magnon excitations are not the origin of the dispersive mode.

%theory
In order to identify the origin of the dispersive mode, we perform theoretical calculations of the dynamical charge structure factor on oxygen orbitals in a three-band Hubbard model for the CuO$_2$ plane with the hopping between Cu3$d_{x^2-y^2}$ and O2$p_\sigma$ orbitals, $T_{pd}$, the hopping between neighboring O2$p_\sigma$ orbitals, $T_{pp}$, the charge transfer energy between Cu3$d_{x^2-y^2}$ and O2$p_\sigma$ orbitals, $\Delta$, the on-site Coulomb interaction on Cu3$d_{x^2-y^2}$, $U_d$, and the on-site Coulomb interactions on O2$p_\sigma$,  $U_p$ (see Supplemental Material S2~\cite{Supple}).
We take a typical parameter set for cuprate superconductors~\cite{Maekawa}: $T_{pd}=1$~eV, $T_{pp}=0.3$~eV, $\Delta=3$~eV, $U_d=8$~eV, and $U_p=4$~eV. 
The dynamical charge structure factor on an orbital $\phi$ is given by
%\begin{equation*}
$N_\phi \left( \mathbf{q},\omega \right) =  \sum_f \left| \left\langle f \right| N^\phi_\mathbf{q} \left| 0 \right\rangle \right|^2 \delta(\omega-E_f+E_0)$,
%\end{equation*}
where $\left| 0 \right\rangle$ and $\left| f \right\rangle$ represent the ground state final state with energy $E_0$ and $E_f$, respectively, and $N^\phi_\mathbf{q}$ is the Fourier-transformed number operator on $\phi$.

We perform a large-scale dynamical density-matrix renormalization-group (DMRG) calculation of $N_\phi \left( \mathbf{q},\omega \right)$ for a small system with $6\times 4=24$ CuO$_2$ units, where a cylindrical geometry with periodic (open) boundary conditions along the $x$ ($y$) direction is introduced.
The numerical method is detailed in the Supplemental Material S2~\cite{Supple}.
We consider $\mathbf{q}$ with minimum $q_y$, i.e., $q_y=1/10$, to make a comparison with the experimental data along the $(h,0)$ direction. 
Since the $\sigma$-polarized incident x-rays are employed in the experiments, the $1s$ core electron is excited to 2$p$ orbitals perpendicular to the $(h,0)$ direction in the RIXS process, indicating predominant polarization for 2$p_y$ orbitals.
Therefore, we take $\phi=2p_y$.

%\bibitem{Supple}See Supplemental Material for the details about theoretical two-magnon Raman scattering and dynamical DMRG.

\begin{figure}[t]
\includegraphics[width=0.4\textwidth]{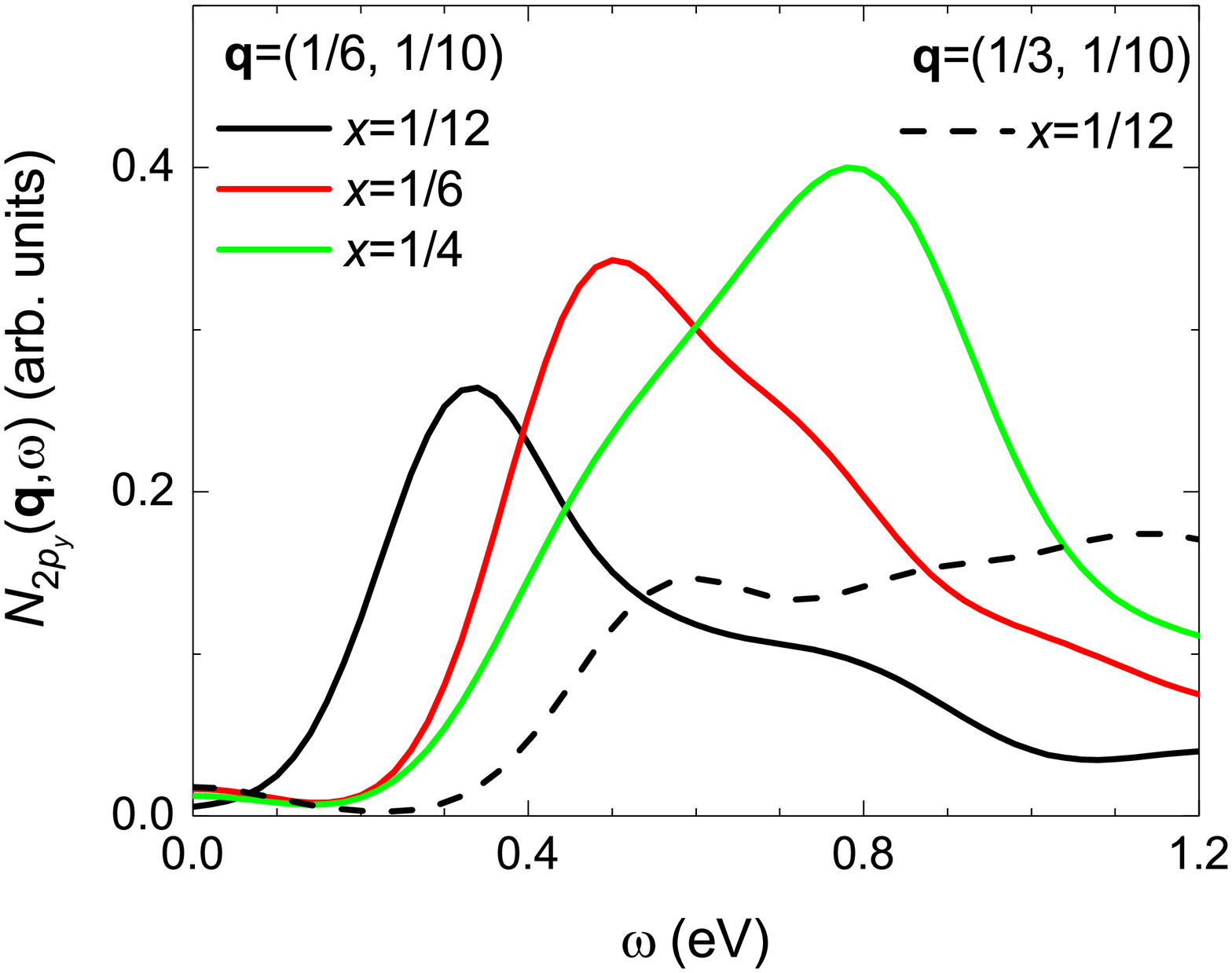}
\caption{
(color online). Dynamical charge structure factor on O2$p_y$ orbital in a $6\times 4$ cylindrical three-band Hubbard cluster with $T_{pd}=1$~eV, $T_{pp}=0.3$~eV, $\Delta=3$~eV, $U_d=8$~eV, and $U_p=4$~eV. Black, red, and green solid lines represent spectra with hole concertation $x=1/12$, $1/6$, and $1/4$, respectively, at $\mathbf{q}=(1/6,1/10)$. Black broken line is for $x=1/12$ at $\mathbf{q}=(1/3,1/10)$. A Gaussian broadening width of 0.1~eV is used for the spectral weights. We note that small weights centered at $\omega=0$~eV are due to less convergence in dynamical DMRG.
}
\label{fig3}
\end{figure}

Figure~\ref{fig3} shows $N_{2p_y} \left( \mathbf{q},\omega \right)$ for hole concentrations $x=1/12$, $1/6$, and $1/4$ at $\mathbf{q}=(1/6,1/10)$ and for $x=1/12$ at $\mathbf{q}=(1/3,1/10)$. The broad peak shifts to higher energy with increasing $x$, accompanied by the increase of spectral weight. The peak positons at $\mathbf{q}=(1/6,1/10)$ (0.3~eV at $x=1/12$, 0.5~eV at $x=1/6$, 0.8~eV at $x=1/4$) are comparable to the experimental data near $\mathbf{q}=(0.16,0)$ (0.49~eV at $x=0.075$ in Fig.~\ref{fig2}(f), 0.69~eV at $x=0.18$ in Fig.~\ref{fig2}(i), 0.65eV at $x=0.25$ in Fig.~\ref{fig2}(j)). With increasing $\mathbf{q}$ from $q_x=1/6$ to $1/3$, the spectral weight shifts to higher energy as expected. Since these momentum and doping dependences of $N_{2p_y} \left( \mathbf{q},\omega \right)$ are consistent with the experimental observation, we ascribe the dispersive mode to the intraband charge excitations.

%discussion
%We have thus found a counterpart of spin excitations in the hole-doped cuprates, namely, charge excitations observed by O $K$-edge RIXS.
We have thus confirmed the charge excitations, which are the counterpart of spin excitations, in the hole-doped cuprates by O $K$-edge RIXS.
While the energy of the spin excitations (paramagnons) in LSCO is at most 0.3 eV \cite{Dean2013,Dean2013c,Wakimoto2015,Monney2016}, the charge excitations are observed up to 0.6-0.8 eV.
These energy ranges of the excitations are reasonable because the magnitude of the dispersion of the spin and charge excitations are respectively scaled by the exchange interaction ($J \sim$ 0.1 eV) and the transfer energy ($t \sim$ 0.4 eV) in the terminology of the $t$-$J$ model.
The intraband charge excitations extend above 1 eV beyond the accessible Brillouin zone of the O $K$-edge and the high-energy part of the excitations has been observed for overdoped compounds by Cu $K$-edge RIXS \cite{Wakimoto2013}.
Slight shift of the peak position with changing $E_i$ (non-perfect Raman behavior) shown in Fig.\ref{fig1}(c) and the broad width of the peak suggest that the experimentally observed dispersive feature is incoherent \cite{Benjamin2014,Tsutsui2016}.
Then the feature comes from particle-hole charge excitations rather than coherent plasmon excitations.
One may connect the charge excitations to the charge order which has recently attracted great interest as a competing phenomenon to superconductivity \cite{Tranquada1995,Ghiringhelli2012,Comin2016}.
While the spin excitations might change slightly across the propagation vector of the charge order \cite{Dean2013c}, we could not find any change of the charge excitations in LBCO ($x$ = 0.125) across the transition temperature of the charge order within the experimental resolution (see Supplemental Material S3~\cite{Supple}).

Combining the present work with the Cu $L_3$-edge RIXS of Nd$_{2-x}$Ce$_x$CuO$_4$ (NCCO) \cite{Ishii2014,Lee2014}, we proved that a dispersive mode in the sub-eV region exists in both hole and electron doped systems.
It strongly supports that the origin of the mode is charge excitations which are common in the electron dynamics of doped Mott insulators.
When compared quantitatively, the magnitude of the dispersion is larger in NCCO than in LSCO (LBCO) as shown in Fig.~\ref{fig2}(g-h).
Naively, the smaller charge-transfer gap of NCCO \cite{Tokura1990} gives larger hopping energy which agrees with the steeper dispersion, but more systematic studies on various cuprate superconductors are necessary to judge whether the sign of the charge of doped carriers is essential for the quantitative difference of the dispersion.

In summary, we have performed O $K$-edge RIXS study of La$_{2-x}$(Br,Sr)$_x$CuO$_4$ and identified the intraband charge excitations in the hole-doped cuprates.
The charge excitations form a positive dispersing mode and the dispersion becomes steeper with increasing the hole concentration.
Theoretical calculation of the dynamical charge structure factor on oxygen orbitals in a three-band Hubbard model agrees with the experimentally observed momentum and doping dependence.
We conclude that identification of the spin and charge dynamics at the respective energy scale of $J$ and $t$ has been completed for both hole- and electron-doped cuprates.

\begin{acknowledgments}
The authors would like to thank H. Yamase for invaluable discussion and M. Dantz for experimental assistance at the ADRESS beamline.
This work was carried out under the Inter-university Cooperative Research Program of the Institute for Materials Research, Tohoku University (Proposal No.\ 16K0048) and the joint research in the Synchrotron Radiation Research Organization and the Institute for Solid State Physics (ISSP), the University of Tokyo.
The synchrotron radiation experiments at the Paul Scherrer Institut were performed at the ADRESS beamline of the Swiss Light Source and those at SPring-8 were carried out at the BL07LSU with the approval of the Japan Synchrotron Radiation Research Institute (JASRI) (Proposal Numbers 2016A7510 and 2015A7484).
Numerical works were supported by a post-K computer project: Creation of new functional devices and high-performance materials to support next-generation industries (CDMSI) and by HPCI Strategic Programs for Innovative Research (SPIRE) (hp160099 and hp160222).
The numerical calculation was carried out at the K Computer and ISSP, the University of Tokyo.
This work was financially supported by JSPS KAKENHI Grant Numbers 25400333, 26287079, 16H02125, and 16H04004.
J.~P. and T.~S. acknowledge financial support through the Dysenos AG by Kabelwerke Brugg AG Holding, Fachhochschule Nordwestschweiz, and the Paul Scherrer Institut.
J.~P. also acknowledges financial support by the Swiss National Science Foundation Early Postdoc.Mobility fellowship project P2FRP2\_171824.
\end{acknowledgments}

\bibliography{lsco-ok}

\begin{thebibliography}{40}%
\makeatletter
\providecommand \@ifxundefined [1]{%
 \@ifx{#1\undefined}
}%
\providecommand \@ifnum [1]{%
 \ifnum #1\expandafter \@firstoftwo
 \else \expandafter \@secondoftwo
 \fi
}%
\providecommand \@ifx [1]{%
 \ifx #1\expandafter \@firstoftwo
 \else \expandafter \@secondoftwo
 \fi
}%
\providecommand \natexlab [1]{#1}%
\providecommand \enquote  [1]{``#1''}%
\providecommand \bibnamefont  [1]{#1}%
\providecommand \bibfnamefont [1]{#1}%
\providecommand \citenamefont [1]{#1}%
\providecommand \href@noop [0]{\@secondoftwo}%
\providecommand \href [0]{\begingroup \@sanitize@url \@href}%
\providecommand \@href[1]{\@@startlink{#1}\@@href}%
\providecommand \@@href[1]{\endgroup#1\@@endlink}%
\providecommand \@sanitize@url [0]{\catcode `\\12\catcode `\$12\catcode
  `\&12\catcode `\#12\catcode `\^12\catcode `\_12\catcode `\%12\relax}%
\providecommand \@@startlink[1]{}%
\providecommand \@@endlink[0]{}%
\providecommand \url  [0]{\begingroup\@sanitize@url \@url }%
\providecommand \@url [1]{\endgroup\@href {#1}{\urlprefix }}%
\providecommand \urlprefix  [0]{URL }%
\providecommand \Eprint [0]{\href }%
\providecommand \doibase [0]{http://dx.doi.org/}%
\providecommand \selectlanguage [0]{\@gobble}%
\providecommand \bibinfo  [0]{\@secondoftwo}%
\providecommand \bibfield  [0]{\@secondoftwo}%
\providecommand \translation [1]{[#1]}%
\providecommand \BibitemOpen [0]{}%
\providecommand \bibitemStop [0]{}%
\providecommand \bibitemNoStop [0]{.\EOS\space}%
\providecommand \EOS [0]{\spacefactor3000\relax}%
\providecommand \BibitemShut  [1]{\csname bibitem#1\endcsname}%
\let\auto@bib@innerbib\@empty
%</preamble>
\bibitem [{\citenamefont {Imada}\ \emph {et~al.}(1998)\citenamefont {Imada},
  \citenamefont {Fujimori},\ and\ \citenamefont {Tokura}}]{Imada1998}%
  \BibitemOpen
  \bibfield  {author} {\bibinfo {author} {\bibfnamefont {M.}~\bibnamefont
  {Imada}}, \bibinfo {author} {\bibfnamefont {A.}~\bibnamefont {Fujimori}}, \
  and\ \bibinfo {author} {\bibfnamefont {Y.}~\bibnamefont {Tokura}},\ }\href
  {\doibase 10.1103/RevModPhys.70.1039} {\bibfield  {journal} {\bibinfo
  {journal} {Rev. Mod. Phys.}\ }\textbf {\bibinfo {volume} {70}},\ \bibinfo
  {pages} {1039} (\bibinfo {year} {1998})}\BibitemShut {NoStop}%
\bibitem [{\citenamefont {Lee}\ \emph {et~al.}(2006)\citenamefont {Lee},
  \citenamefont {Nagaosa},\ and\ \citenamefont {Wen}}]{Lee2006}%
  \BibitemOpen
  \bibfield  {author} {\bibinfo {author} {\bibfnamefont {P.~A.}\ \bibnamefont
  {Lee}}, \bibinfo {author} {\bibfnamefont {N.}~\bibnamefont {Nagaosa}}, \ and\
  \bibinfo {author} {\bibfnamefont {X.-G.}\ \bibnamefont {Wen}},\ }\href
  {\doibase 10.1103/RevModPhys.78.17} {\bibfield  {journal} {\bibinfo
  {journal} {Rev. Mod. Phys.}\ }\textbf {\bibinfo {volume} {78}},\ \bibinfo
  {pages} {17} (\bibinfo {year} {2006})}\BibitemShut {NoStop}%
\bibitem [{\citenamefont {Keimer}\ \emph {et~al.}(2015)\citenamefont {Keimer},
  \citenamefont {Kivelson}, \citenamefont {Norman}, \citenamefont {Uchida},\
  and\ \citenamefont {Zaanen}}]{Keimer2015}%
  \BibitemOpen
  \bibfield  {author} {\bibinfo {author} {\bibfnamefont {B.}~\bibnamefont
  {Keimer}}, \bibinfo {author} {\bibfnamefont {S.~A.}\ \bibnamefont
  {Kivelson}}, \bibinfo {author} {\bibfnamefont {M.~R.}\ \bibnamefont
  {Norman}}, \bibinfo {author} {\bibfnamefont {S.}~\bibnamefont {Uchida}}, \
  and\ \bibinfo {author} {\bibfnamefont {J.}~\bibnamefont {Zaanen}},\ }\href
  {http://dx.doi.org/10.1038/nature14165} {\bibfield  {journal} {\bibinfo
  {journal} {Nature}\ }\textbf {\bibinfo {volume} {518}},\ \bibinfo {pages}
  {179} (\bibinfo {year} {2015})}\BibitemShut {NoStop}%
\bibitem [{\citenamefont {Ament}\ \emph {et~al.}(2009)\citenamefont {Ament},
  \citenamefont {Ghiringhelli}, \citenamefont {Sala}, \citenamefont
  {Braicovich},\ and\ \citenamefont {van~den Brink}}]{Ament2009b}%
  \BibitemOpen
  \bibfield  {author} {\bibinfo {author} {\bibfnamefont {L.~J.~P.}\
  \bibnamefont {Ament}}, \bibinfo {author} {\bibfnamefont {G.}~\bibnamefont
  {Ghiringhelli}}, \bibinfo {author} {\bibfnamefont {M.~M.}\ \bibnamefont
  {Sala}}, \bibinfo {author} {\bibfnamefont {L.}~\bibnamefont {Braicovich}}, \
  and\ \bibinfo {author} {\bibfnamefont {J.}~\bibnamefont {van~den Brink}},\
  }\href {\doibase 10.1103/PhysRevLett.103.117003} {\bibfield  {journal}
  {\bibinfo  {journal} {Phys. Rev. Lett.}\ }\textbf {\bibinfo {volume} {103}},\
  \bibinfo {eid} {117003} (\bibinfo {year} {2009})}\BibitemShut {NoStop}%
\bibitem [{\citenamefont {Braicovich}\ \emph {et~al.}(2010)\citenamefont
  {Braicovich}, \citenamefont {van~den Brink}, \citenamefont {Bisogni},
  \citenamefont {Moretti~Sala}, \citenamefont {Ament}, \citenamefont {Brookes},
  \citenamefont {De~Luca}, \citenamefont {Salluzzo}, \citenamefont {Schmitt},
  \citenamefont {Strocov},\ and\ \citenamefont
  {Ghiringhelli}}]{Braicovich2010}%
  \BibitemOpen
  \bibfield  {author} {\bibinfo {author} {\bibfnamefont {L.}~\bibnamefont
  {Braicovich}}, \bibinfo {author} {\bibfnamefont {J.}~\bibnamefont {van~den
  Brink}}, \bibinfo {author} {\bibfnamefont {V.}~\bibnamefont {Bisogni}},
  \bibinfo {author} {\bibfnamefont {M.}~\bibnamefont {Moretti~Sala}}, \bibinfo
  {author} {\bibfnamefont {L.~J.~P.}\ \bibnamefont {Ament}}, \bibinfo {author}
  {\bibfnamefont {N.~B.}\ \bibnamefont {Brookes}}, \bibinfo {author}
  {\bibfnamefont {G.~M.}\ \bibnamefont {De~Luca}}, \bibinfo {author}
  {\bibfnamefont {M.}~\bibnamefont {Salluzzo}}, \bibinfo {author}
  {\bibfnamefont {T.}~\bibnamefont {Schmitt}}, \bibinfo {author} {\bibfnamefont
  {V.~N.}\ \bibnamefont {Strocov}}, \ and\ \bibinfo {author} {\bibfnamefont
  {G.}~\bibnamefont {Ghiringhelli}},\ }\href {\doibase
  10.1103/PhysRevLett.104.077002} {\bibfield  {journal} {\bibinfo  {journal}
  {Phys. Rev. Lett.}\ }\textbf {\bibinfo {volume} {104}},\ \bibinfo {pages}
  {077002} (\bibinfo {year} {2010})}\BibitemShut {NoStop}%
\bibitem [{\citenamefont {Uchida}\ \emph {et~al.}(1991)\citenamefont {Uchida},
  \citenamefont {Ido}, \citenamefont {Takagi}, \citenamefont {Arima},
  \citenamefont {Tokura},\ and\ \citenamefont {Tajima}}]{Uchida1991}%
  \BibitemOpen
  \bibfield  {author} {\bibinfo {author} {\bibfnamefont {S.}~\bibnamefont
  {Uchida}}, \bibinfo {author} {\bibfnamefont {T.}~\bibnamefont {Ido}},
  \bibinfo {author} {\bibfnamefont {H.}~\bibnamefont {Takagi}}, \bibinfo
  {author} {\bibfnamefont {T.}~\bibnamefont {Arima}}, \bibinfo {author}
  {\bibfnamefont {Y.}~\bibnamefont {Tokura}}, \ and\ \bibinfo {author}
  {\bibfnamefont {S.}~\bibnamefont {Tajima}},\ }\href {\doibase
  10.1103/PhysRevB.43.7942} {\bibfield  {journal} {\bibinfo  {journal} {Phys.
  Rev. B}\ }\textbf {\bibinfo {volume} {43}},\ \bibinfo {pages} {7942}
  (\bibinfo {year} {1991})}\BibitemShut {NoStop}%
\bibitem [{\citenamefont {Onose}\ \emph {et~al.}(2004)\citenamefont {Onose},
  \citenamefont {Taguchi}, \citenamefont {Ishizaka},\ and\ \citenamefont
  {Tokura}}]{Onose2004}%
  \BibitemOpen
  \bibfield  {author} {\bibinfo {author} {\bibfnamefont {Y.}~\bibnamefont
  {Onose}}, \bibinfo {author} {\bibfnamefont {Y.}~\bibnamefont {Taguchi}},
  \bibinfo {author} {\bibfnamefont {K.}~\bibnamefont {Ishizaka}}, \ and\
  \bibinfo {author} {\bibfnamefont {Y.}~\bibnamefont {Tokura}},\ }\href
  {\doibase 10.1103/PhysRevB.69.024504} {\bibfield  {journal} {\bibinfo
  {journal} {Phys. Rev. B}\ }\textbf {\bibinfo {volume} {69}},\ \bibinfo
  {pages} {024504} (\bibinfo {year} {2004})}\BibitemShut {NoStop}%
\bibitem [{\citenamefont {Kim}\ \emph {et~al.}(2004)\citenamefont {Kim},
  \citenamefont {Hill}, \citenamefont {Komiya}, \citenamefont {Ando},
  \citenamefont {Casa}, \citenamefont {Gog},\ and\ \citenamefont
  {Venkataraman}}]{Kim2004c}%
  \BibitemOpen
  \bibfield  {author} {\bibinfo {author} {\bibfnamefont {Y.-J.}\ \bibnamefont
  {Kim}}, \bibinfo {author} {\bibfnamefont {J.~P.}\ \bibnamefont {Hill}},
  \bibinfo {author} {\bibfnamefont {S.}~\bibnamefont {Komiya}}, \bibinfo
  {author} {\bibfnamefont {Y.}~\bibnamefont {Ando}}, \bibinfo {author}
  {\bibfnamefont {D.}~\bibnamefont {Casa}}, \bibinfo {author} {\bibfnamefont
  {T.}~\bibnamefont {Gog}}, \ and\ \bibinfo {author} {\bibfnamefont {C.~T.}\
  \bibnamefont {Venkataraman}},\ }\href {\doibase 10.1103/PhysRevB.70.094524}
  {\bibfield  {journal} {\bibinfo  {journal} {Phys. Rev. B}\ }\textbf {\bibinfo
  {volume} {70}},\ \bibinfo {eid} {094524} (\bibinfo {year}
  {2004})}\BibitemShut {NoStop}%
\bibitem [{\citenamefont {Ishii}\ \emph {et~al.}(2005)\citenamefont {Ishii},
  \citenamefont {Tsutsui}, \citenamefont {Endoh}, \citenamefont {Tohyama},
  \citenamefont {Maekawa}, \citenamefont {Hoesch}, \citenamefont {Kuzushita},
  \citenamefont {Tsubota}, \citenamefont {Inami}, \citenamefont {Mizuki},
  \citenamefont {Murakami},\ and\ \citenamefont {Yamada}}]{Ishii2005b}%
  \BibitemOpen
  \bibfield  {author} {\bibinfo {author} {\bibfnamefont {K.}~\bibnamefont
  {Ishii}}, \bibinfo {author} {\bibfnamefont {K.}~\bibnamefont {Tsutsui}},
  \bibinfo {author} {\bibfnamefont {Y.}~\bibnamefont {Endoh}}, \bibinfo
  {author} {\bibfnamefont {T.}~\bibnamefont {Tohyama}}, \bibinfo {author}
  {\bibfnamefont {S.}~\bibnamefont {Maekawa}}, \bibinfo {author} {\bibfnamefont
  {M.}~\bibnamefont {Hoesch}}, \bibinfo {author} {\bibfnamefont
  {K.}~\bibnamefont {Kuzushita}}, \bibinfo {author} {\bibfnamefont
  {M.}~\bibnamefont {Tsubota}}, \bibinfo {author} {\bibfnamefont
  {T.}~\bibnamefont {Inami}}, \bibinfo {author} {\bibfnamefont
  {J.}~\bibnamefont {Mizuki}}, \bibinfo {author} {\bibfnamefont
  {Y.}~\bibnamefont {Murakami}}, \ and\ \bibinfo {author} {\bibfnamefont
  {K.}~\bibnamefont {Yamada}},\ }\href {\doibase 10.1103/PhysRevLett.94.207003}
  {\bibfield  {journal} {\bibinfo  {journal} {Phys. Rev. Lett.}\ }\textbf
  {\bibinfo {volume} {94}},\ \bibinfo {eid} {207003} (\bibinfo {year}
  {2005})}\BibitemShut {NoStop}%
\bibitem [{\citenamefont {Wakimoto}\ \emph {et~al.}(2013)\citenamefont
  {Wakimoto}, \citenamefont {Ishii}, \citenamefont {Kimura}, \citenamefont
  {Ikeuchi}, \citenamefont {Yoshida}, \citenamefont {Adachi}, \citenamefont
  {Casa}, \citenamefont {Fujita}, \citenamefont {Fukunaga}, \citenamefont
  {Gog}, \citenamefont {Koike}, \citenamefont {Mizuki},\ and\ \citenamefont
  {Yamada}}]{Wakimoto2013}%
  \BibitemOpen
  \bibfield  {author} {\bibinfo {author} {\bibfnamefont {S.}~\bibnamefont
  {Wakimoto}}, \bibinfo {author} {\bibfnamefont {K.}~\bibnamefont {Ishii}},
  \bibinfo {author} {\bibfnamefont {H.}~\bibnamefont {Kimura}}, \bibinfo
  {author} {\bibfnamefont {K.}~\bibnamefont {Ikeuchi}}, \bibinfo {author}
  {\bibfnamefont {M.}~\bibnamefont {Yoshida}}, \bibinfo {author} {\bibfnamefont
  {T.}~\bibnamefont {Adachi}}, \bibinfo {author} {\bibfnamefont
  {D.}~\bibnamefont {Casa}}, \bibinfo {author} {\bibfnamefont {M.}~\bibnamefont
  {Fujita}}, \bibinfo {author} {\bibfnamefont {Y.}~\bibnamefont {Fukunaga}},
  \bibinfo {author} {\bibfnamefont {T.}~\bibnamefont {Gog}}, \bibinfo {author}
  {\bibfnamefont {Y.}~\bibnamefont {Koike}}, \bibinfo {author} {\bibfnamefont
  {J.}~\bibnamefont {Mizuki}}, \ and\ \bibinfo {author} {\bibfnamefont
  {K.}~\bibnamefont {Yamada}},\ }\href {\doibase 10.1103/PhysRevB.87.104511}
  {\bibfield  {journal} {\bibinfo  {journal} {Phys. Rev. B}\ }\textbf {\bibinfo
  {volume} {87}},\ \bibinfo {pages} {104511} (\bibinfo {year}
  {2013})}\BibitemShut {NoStop}%
\bibitem [{\citenamefont {Jia}\ \emph {et~al.}(2016)\citenamefont {Jia},
  \citenamefont {Wohlfeld}, \citenamefont {Wang}, \citenamefont {Moritz},\ and\
  \citenamefont {Devereaux}}]{Jia2016}%
  \BibitemOpen
  \bibfield  {author} {\bibinfo {author} {\bibfnamefont {C.}~\bibnamefont
  {Jia}}, \bibinfo {author} {\bibfnamefont {K.}~\bibnamefont {Wohlfeld}},
  \bibinfo {author} {\bibfnamefont {Y.}~\bibnamefont {Wang}}, \bibinfo {author}
  {\bibfnamefont {B.}~\bibnamefont {Moritz}}, \ and\ \bibinfo {author}
  {\bibfnamefont {T.~P.}\ \bibnamefont {Devereaux}},\ }\href {\doibase
  10.1103/PhysRevX.6.021020} {\bibfield  {journal} {\bibinfo  {journal} {Phys.
  Rev. X}\ }\textbf {\bibinfo {volume} {6}},\ \bibinfo {pages} {021020}
  (\bibinfo {year} {2016})}\BibitemShut {NoStop}%
\bibitem [{\citenamefont {Tsutsui}\ and\ \citenamefont
  {Tohyama}(2016)}]{Tsutsui2016}%
  \BibitemOpen
  \bibfield  {author} {\bibinfo {author} {\bibfnamefont {K.}~\bibnamefont
  {Tsutsui}}\ and\ \bibinfo {author} {\bibfnamefont {T.}~\bibnamefont
  {Tohyama}},\ }\href {\doibase 10.1103/PhysRevB.94.085144} {\bibfield
  {journal} {\bibinfo  {journal} {Phys. Rev. B}\ }\textbf {\bibinfo {volume}
  {94}},\ \bibinfo {pages} {085144} (\bibinfo {year} {2016})}\BibitemShut
  {NoStop}%
\bibitem [{\citenamefont {Monney}\ \emph {et~al.}(2012)\citenamefont {Monney},
  \citenamefont {Zhou}, \citenamefont {Cercellier}, \citenamefont {Vydrova},
  \citenamefont {Garnier}, \citenamefont {Monney}, \citenamefont {Strocov},
  \citenamefont {Berger}, \citenamefont {Beck}, \citenamefont {Schmitt},\ and\
  \citenamefont {Aebi}}]{Monney2012}%
  \BibitemOpen
  \bibfield  {author} {\bibinfo {author} {\bibfnamefont {C.}~\bibnamefont
  {Monney}}, \bibinfo {author} {\bibfnamefont {K.~J.}\ \bibnamefont {Zhou}},
  \bibinfo {author} {\bibfnamefont {H.}~\bibnamefont {Cercellier}}, \bibinfo
  {author} {\bibfnamefont {Z.}~\bibnamefont {Vydrova}}, \bibinfo {author}
  {\bibfnamefont {M.~G.}\ \bibnamefont {Garnier}}, \bibinfo {author}
  {\bibfnamefont {G.}~\bibnamefont {Monney}}, \bibinfo {author} {\bibfnamefont
  {V.~N.}\ \bibnamefont {Strocov}}, \bibinfo {author} {\bibfnamefont
  {H.}~\bibnamefont {Berger}}, \bibinfo {author} {\bibfnamefont
  {H.}~\bibnamefont {Beck}}, \bibinfo {author} {\bibfnamefont {T.}~\bibnamefont
  {Schmitt}}, \ and\ \bibinfo {author} {\bibfnamefont {P.}~\bibnamefont
  {Aebi}},\ }\href {\doibase 10.1103/PhysRevLett.109.047401} {\bibfield
  {journal} {\bibinfo  {journal} {Phys. Rev. Lett.}\ }\textbf {\bibinfo
  {volume} {109}},\ \bibinfo {pages} {047401} (\bibinfo {year}
  {2012})}\BibitemShut {NoStop}%
\bibitem [{\citenamefont {Ishii}\ \emph {et~al.}(2014)\citenamefont {Ishii},
  \citenamefont {Fujita}, \citenamefont {Sasaki}, \citenamefont {Minola},
  \citenamefont {Dellea}, \citenamefont {Mazzoli}, \citenamefont {Kummer},
  \citenamefont {Ghiringhelli}, \citenamefont {Braicovich}, \citenamefont
  {Tohyama}, \citenamefont {Tsutsumi}, \citenamefont {Sato}, \citenamefont
  {Kajimoto}, \citenamefont {Ikeuchi}, \citenamefont {Yamada}, \citenamefont
  {Yoshida}, \citenamefont {Kurooka},\ and\ \citenamefont
  {Mizuki}}]{Ishii2014}%
  \BibitemOpen
  \bibfield  {author} {\bibinfo {author} {\bibfnamefont {K.}~\bibnamefont
  {Ishii}}, \bibinfo {author} {\bibfnamefont {M.}~\bibnamefont {Fujita}},
  \bibinfo {author} {\bibfnamefont {T.}~\bibnamefont {Sasaki}}, \bibinfo
  {author} {\bibfnamefont {M.}~\bibnamefont {Minola}}, \bibinfo {author}
  {\bibfnamefont {G.}~\bibnamefont {Dellea}}, \bibinfo {author} {\bibfnamefont
  {C.}~\bibnamefont {Mazzoli}}, \bibinfo {author} {\bibfnamefont
  {K.}~\bibnamefont {Kummer}}, \bibinfo {author} {\bibfnamefont
  {G.}~\bibnamefont {Ghiringhelli}}, \bibinfo {author} {\bibfnamefont
  {L.}~\bibnamefont {Braicovich}}, \bibinfo {author} {\bibfnamefont
  {T.}~\bibnamefont {Tohyama}}, \bibinfo {author} {\bibfnamefont
  {K.}~\bibnamefont {Tsutsumi}}, \bibinfo {author} {\bibfnamefont
  {K.}~\bibnamefont {Sato}}, \bibinfo {author} {\bibfnamefont {R.}~\bibnamefont
  {Kajimoto}}, \bibinfo {author} {\bibfnamefont {K.}~\bibnamefont {Ikeuchi}},
  \bibinfo {author} {\bibfnamefont {K.}~\bibnamefont {Yamada}}, \bibinfo
  {author} {\bibfnamefont {M.}~\bibnamefont {Yoshida}}, \bibinfo {author}
  {\bibfnamefont {M.}~\bibnamefont {Kurooka}}, \ and\ \bibinfo {author}
  {\bibfnamefont {J.}~\bibnamefont {Mizuki}},\ }\href
  {http://dx.doi.org/10.1038/ncomms4714} {\bibfield  {journal} {\bibinfo
  {journal} {Nat. Commun.}\ }\textbf {\bibinfo {volume} {5}},\ \bibinfo {pages}
  {3714} (\bibinfo {year} {2014})}\BibitemShut {NoStop}%
\bibitem [{\citenamefont {Lee}\ \emph {et~al.}(2014)\citenamefont {Lee},
  \citenamefont {Lee}, \citenamefont {Nowadnick}, \citenamefont {Gerber},
  \citenamefont {Tabis}, \citenamefont {Huang}, \citenamefont {Strocov},
  \citenamefont {Motoyama}, \citenamefont {Yu}, \citenamefont {Moritz},
  \citenamefont {Huang}, \citenamefont {Wang}, \citenamefont {Huang},
  \citenamefont {Wu}, \citenamefont {Chen}, \citenamefont {Huang},
  \citenamefont {Greven}, \citenamefont {Schmitt}, \citenamefont {Shen},\ and\
  \citenamefont {Devereaux}}]{Lee2014}%
  \BibitemOpen
  \bibfield  {author} {\bibinfo {author} {\bibfnamefont {W.~S.}\ \bibnamefont
  {Lee}}, \bibinfo {author} {\bibfnamefont {J.~J.}\ \bibnamefont {Lee}},
  \bibinfo {author} {\bibfnamefont {E.~A.}\ \bibnamefont {Nowadnick}}, \bibinfo
  {author} {\bibfnamefont {S.}~\bibnamefont {Gerber}}, \bibinfo {author}
  {\bibfnamefont {W.}~\bibnamefont {Tabis}}, \bibinfo {author} {\bibfnamefont
  {S.~W.}\ \bibnamefont {Huang}}, \bibinfo {author} {\bibfnamefont {V.~N.}\
  \bibnamefont {Strocov}}, \bibinfo {author} {\bibfnamefont {E.~M.}\
  \bibnamefont {Motoyama}}, \bibinfo {author} {\bibfnamefont {G.}~\bibnamefont
  {Yu}}, \bibinfo {author} {\bibfnamefont {B.}~\bibnamefont {Moritz}}, \bibinfo
  {author} {\bibfnamefont {H.~Y.}\ \bibnamefont {Huang}}, \bibinfo {author}
  {\bibfnamefont {R.~P.}\ \bibnamefont {Wang}}, \bibinfo {author}
  {\bibfnamefont {Y.~B.}\ \bibnamefont {Huang}}, \bibinfo {author}
  {\bibfnamefont {W.~B.}\ \bibnamefont {Wu}}, \bibinfo {author} {\bibfnamefont
  {C.~T.}\ \bibnamefont {Chen}}, \bibinfo {author} {\bibfnamefont {D.~J.}\
  \bibnamefont {Huang}}, \bibinfo {author} {\bibfnamefont {M.}~\bibnamefont
  {Greven}}, \bibinfo {author} {\bibfnamefont {T.}~\bibnamefont {Schmitt}},
  \bibinfo {author} {\bibfnamefont {Z.~X.}\ \bibnamefont {Shen}}, \ and\
  \bibinfo {author} {\bibfnamefont {T.~P.}\ \bibnamefont {Devereaux}},\ }\href
  {http://dx.doi.org/10.1038/nphys3117} {\bibfield  {journal} {\bibinfo
  {journal} {Nat. Phys.}\ }\textbf {\bibinfo {volume} {10}},\ \bibinfo {pages}
  {883} (\bibinfo {year} {2014})}\BibitemShut {NoStop}%
\bibitem [{\citenamefont {Greco}\ \emph {et~al.}(2016)\citenamefont {Greco},
  \citenamefont {Yamase},\ and\ \citenamefont {Bejas}}]{Greco2016}%
  \BibitemOpen
  \bibfield  {author} {\bibinfo {author} {\bibfnamefont {A.}~\bibnamefont
  {Greco}}, \bibinfo {author} {\bibfnamefont {H.}~\bibnamefont {Yamase}}, \
  and\ \bibinfo {author} {\bibfnamefont {M.}~\bibnamefont {Bejas}},\ }\href
  {\doibase 10.1103/PhysRevB.94.075139} {\bibfield  {journal} {\bibinfo
  {journal} {Phys. Rev. B}\ }\textbf {\bibinfo {volume} {94}},\ \bibinfo
  {pages} {075139} (\bibinfo {year} {2016})}\BibitemShut {NoStop}%
\bibitem [{\citenamefont {Ghiringhelli}\ \emph {et~al.}(2006)\citenamefont
  {Ghiringhelli}, \citenamefont {Piazzalunga}, \citenamefont {Dallera},
  \citenamefont {Trezzi}, \citenamefont {Braicovich}, \citenamefont {Schmitt},
  \citenamefont {Strocov}, \citenamefont {Betemps}, \citenamefont {Patthey},
  \citenamefont {Wang},\ and\ \citenamefont {Grioni}}]{Ghiringhelli2006}%
  \BibitemOpen
  \bibfield  {author} {\bibinfo {author} {\bibfnamefont {G.}~\bibnamefont
  {Ghiringhelli}}, \bibinfo {author} {\bibfnamefont {A.}~\bibnamefont
  {Piazzalunga}}, \bibinfo {author} {\bibfnamefont {C.}~\bibnamefont
  {Dallera}}, \bibinfo {author} {\bibfnamefont {G.}~\bibnamefont {Trezzi}},
  \bibinfo {author} {\bibfnamefont {L.}~\bibnamefont {Braicovich}}, \bibinfo
  {author} {\bibfnamefont {T.}~\bibnamefont {Schmitt}}, \bibinfo {author}
  {\bibfnamefont {V.~N.}\ \bibnamefont {Strocov}}, \bibinfo {author}
  {\bibfnamefont {R.}~\bibnamefont {Betemps}}, \bibinfo {author} {\bibfnamefont
  {L.}~\bibnamefont {Patthey}}, \bibinfo {author} {\bibfnamefont
  {X.}~\bibnamefont {Wang}}, \ and\ \bibinfo {author} {\bibfnamefont
  {M.}~\bibnamefont {Grioni}},\ }\href {\doibase 10.1063/1.2372731} {\bibfield
  {journal} {\bibinfo  {journal} {Rev. Sci. Instrum.}\ }\textbf {\bibinfo
  {volume} {77}},\ \bibinfo {eid} {113108} (\bibinfo {year}
  {2006})}\BibitemShut {NoStop}%
\bibitem [{\citenamefont {Strocov}\ \emph {et~al.}(2010)\citenamefont
  {Strocov}, \citenamefont {Schmitt}, \citenamefont {Flechsig}, \citenamefont
  {Schmidt}, \citenamefont {Imhof}, \citenamefont {Chen}, \citenamefont
  {Raabe}, \citenamefont {Betemps}, \citenamefont {Zimoch}, \citenamefont
  {Krempasky}, \citenamefont {Wang}, \citenamefont {Grioni}, \citenamefont
  {Piazzalunga},\ and\ \citenamefont {Patthey}}]{Strocov2010}%
  \BibitemOpen
  \bibfield  {author} {\bibinfo {author} {\bibfnamefont {V.~N.}\ \bibnamefont
  {Strocov}}, \bibinfo {author} {\bibfnamefont {T.}~\bibnamefont {Schmitt}},
  \bibinfo {author} {\bibfnamefont {U.}~\bibnamefont {Flechsig}}, \bibinfo
  {author} {\bibfnamefont {T.}~\bibnamefont {Schmidt}}, \bibinfo {author}
  {\bibfnamefont {A.}~\bibnamefont {Imhof}}, \bibinfo {author} {\bibfnamefont
  {Q.}~\bibnamefont {Chen}}, \bibinfo {author} {\bibfnamefont {J.}~\bibnamefont
  {Raabe}}, \bibinfo {author} {\bibfnamefont {R.}~\bibnamefont {Betemps}},
  \bibinfo {author} {\bibfnamefont {D.}~\bibnamefont {Zimoch}}, \bibinfo
  {author} {\bibfnamefont {J.}~\bibnamefont {Krempasky}}, \bibinfo {author}
  {\bibfnamefont {X.}~\bibnamefont {Wang}}, \bibinfo {author} {\bibfnamefont
  {M.}~\bibnamefont {Grioni}}, \bibinfo {author} {\bibfnamefont
  {A.}~\bibnamefont {Piazzalunga}}, \ and\ \bibinfo {author} {\bibfnamefont
  {L.}~\bibnamefont {Patthey}},\ }\href {\doibase 10.1107/S0909049510019862}
  {\bibfield  {journal} {\bibinfo  {journal} {J. Synchrotron Radiat.}\ }\textbf
  {\bibinfo {volume} {17}},\ \bibinfo {pages} {631} (\bibinfo {year}
  {2010})}\BibitemShut {NoStop}%
\bibitem [{\citenamefont {Harada}\ \emph {et~al.}(2012)\citenamefont {Harada},
  \citenamefont {Kobayashi}, \citenamefont {Niwa}, \citenamefont {Senba},
  \citenamefont {Ohashi}, \citenamefont {Tokushima}, \citenamefont {Horikawa},
  \citenamefont {Shin},\ and\ \citenamefont {Oshima}}]{Harada2012}%
  \BibitemOpen
  \bibfield  {author} {\bibinfo {author} {\bibfnamefont {Y.}~\bibnamefont
  {Harada}}, \bibinfo {author} {\bibfnamefont {M.}~\bibnamefont {Kobayashi}},
  \bibinfo {author} {\bibfnamefont {H.}~\bibnamefont {Niwa}}, \bibinfo {author}
  {\bibfnamefont {Y.}~\bibnamefont {Senba}}, \bibinfo {author} {\bibfnamefont
  {H.}~\bibnamefont {Ohashi}}, \bibinfo {author} {\bibfnamefont
  {T.}~\bibnamefont {Tokushima}}, \bibinfo {author} {\bibfnamefont
  {Y.}~\bibnamefont {Horikawa}}, \bibinfo {author} {\bibfnamefont
  {S.}~\bibnamefont {Shin}}, \ and\ \bibinfo {author} {\bibfnamefont
  {M.}~\bibnamefont {Oshima}},\ }\href {\doibase 10.1063/1.3680559} {\bibfield
  {journal} {\bibinfo  {journal} {Rev. Sci. Instrum.}\ }\textbf {\bibinfo
  {volume} {83}},\ \bibinfo {eid} {013116} (\bibinfo {year}
  {2012})}\BibitemShut {NoStop}%
\bibitem [{\citenamefont {Senba}\ \emph {et~al.}(2011)\citenamefont {Senba},
  \citenamefont {Yamamoto}, \citenamefont {Ohashi}, \citenamefont {Matsuda},
  \citenamefont {Fujisawa}, \citenamefont {Harasawa}, \citenamefont {Okuda},
  \citenamefont {Takahashi}, \citenamefont {Nariyama}, \citenamefont
  {Matsushita}, \citenamefont {Ohata}, \citenamefont {Furukawa}, \citenamefont
  {Tanaka}, \citenamefont {Takeshita}, \citenamefont {Goto}, \citenamefont
  {Kitamura}, \citenamefont {Kakizaki},\ and\ \citenamefont
  {Oshima}}]{Senba2011}%
  \BibitemOpen
  \bibfield  {author} {\bibinfo {author} {\bibfnamefont {Y.}~\bibnamefont
  {Senba}}, \bibinfo {author} {\bibfnamefont {S.}~\bibnamefont {Yamamoto}},
  \bibinfo {author} {\bibfnamefont {H.}~\bibnamefont {Ohashi}}, \bibinfo
  {author} {\bibfnamefont {I.}~\bibnamefont {Matsuda}}, \bibinfo {author}
  {\bibfnamefont {M.}~\bibnamefont {Fujisawa}}, \bibinfo {author}
  {\bibfnamefont {A.}~\bibnamefont {Harasawa}}, \bibinfo {author}
  {\bibfnamefont {T.}~\bibnamefont {Okuda}}, \bibinfo {author} {\bibfnamefont
  {S.}~\bibnamefont {Takahashi}}, \bibinfo {author} {\bibfnamefont
  {N.}~\bibnamefont {Nariyama}}, \bibinfo {author} {\bibfnamefont
  {T.}~\bibnamefont {Matsushita}}, \bibinfo {author} {\bibfnamefont
  {T.}~\bibnamefont {Ohata}}, \bibinfo {author} {\bibfnamefont
  {Y.}~\bibnamefont {Furukawa}}, \bibinfo {author} {\bibfnamefont
  {T.}~\bibnamefont {Tanaka}}, \bibinfo {author} {\bibfnamefont
  {K.}~\bibnamefont {Takeshita}}, \bibinfo {author} {\bibfnamefont
  {S.}~\bibnamefont {Goto}}, \bibinfo {author} {\bibfnamefont {H.}~\bibnamefont
  {Kitamura}}, \bibinfo {author} {\bibfnamefont {A.}~\bibnamefont {Kakizaki}},
  \ and\ \bibinfo {author} {\bibfnamefont {M.}~\bibnamefont {Oshima}},\ }\href
  {http://www.sciencedirect.com/science/article/pii/S0168900211000118}
  {\bibfield  {journal} {\bibinfo  {journal} {Nucl. Instrum. Methods Phys. Res.
  A}\ }\textbf {\bibinfo {volume} {649}},\ \bibinfo {pages} {58} (\bibinfo
  {year} {2011})}\BibitemShut {NoStop}%
\bibitem [{\citenamefont {Chen}\ \emph {et~al.}(1991)\citenamefont {Chen},
  \citenamefont {Sette}, \citenamefont {Ma}, \citenamefont {Hybertsen},
  \citenamefont {Stechel}, \citenamefont {Foulkes}, \citenamefont {Schulter},
  \citenamefont {Cheong}, \citenamefont {Cooper}, \citenamefont {Rupp},
  \citenamefont {Batlogg}, \citenamefont {Soo}, \citenamefont {Ming},
  \citenamefont {Krol},\ and\ \citenamefont {Kao}}]{Chen1991}%
  \BibitemOpen
  \bibfield  {author} {\bibinfo {author} {\bibfnamefont {C.~T.}\ \bibnamefont
  {Chen}}, \bibinfo {author} {\bibfnamefont {F.}~\bibnamefont {Sette}},
  \bibinfo {author} {\bibfnamefont {Y.}~\bibnamefont {Ma}}, \bibinfo {author}
  {\bibfnamefont {M.~S.}\ \bibnamefont {Hybertsen}}, \bibinfo {author}
  {\bibfnamefont {E.~B.}\ \bibnamefont {Stechel}}, \bibinfo {author}
  {\bibfnamefont {W.~M.~C.}\ \bibnamefont {Foulkes}}, \bibinfo {author}
  {\bibfnamefont {M.}~\bibnamefont {Schulter}}, \bibinfo {author}
  {\bibfnamefont {S.-W.}\ \bibnamefont {Cheong}}, \bibinfo {author}
  {\bibfnamefont {A.~S.}\ \bibnamefont {Cooper}}, \bibinfo {author}
  {\bibfnamefont {L.~W.}\ \bibnamefont {Rupp}}, \bibinfo {author}
  {\bibfnamefont {B.}~\bibnamefont {Batlogg}}, \bibinfo {author} {\bibfnamefont
  {Y.~L.}\ \bibnamefont {Soo}}, \bibinfo {author} {\bibfnamefont {Z.~H.}\
  \bibnamefont {Ming}}, \bibinfo {author} {\bibfnamefont {A.}~\bibnamefont
  {Krol}}, \ and\ \bibinfo {author} {\bibfnamefont {Y.~H.}\ \bibnamefont
  {Kao}},\ }\href {\doibase 10.1103/PhysRevLett.66.104} {\bibfield  {journal}
  {\bibinfo  {journal} {Phys. Rev. Lett.}\ }\textbf {\bibinfo {volume} {66}},\
  \bibinfo {pages} {104} (\bibinfo {year} {1991})}\BibitemShut {NoStop}%
\bibitem [{\citenamefont {Harada}\ \emph {et~al.}(2002)\citenamefont {Harada},
  \citenamefont {Okada}, \citenamefont {Eguchi}, \citenamefont {Kotani},
  \citenamefont {Takagi}, \citenamefont {Takeuchi},\ and\ \citenamefont
  {Shin}}]{Harada2002}%
  \BibitemOpen
  \bibfield  {author} {\bibinfo {author} {\bibfnamefont {Y.}~\bibnamefont
  {Harada}}, \bibinfo {author} {\bibfnamefont {K.}~\bibnamefont {Okada}},
  \bibinfo {author} {\bibfnamefont {R.}~\bibnamefont {Eguchi}}, \bibinfo
  {author} {\bibfnamefont {A.}~\bibnamefont {Kotani}}, \bibinfo {author}
  {\bibfnamefont {H.}~\bibnamefont {Takagi}}, \bibinfo {author} {\bibfnamefont
  {T.}~\bibnamefont {Takeuchi}}, \ and\ \bibinfo {author} {\bibfnamefont
  {S.}~\bibnamefont {Shin}},\ }\href {\doibase 10.1103/PhysRevB.66.165104}
  {\bibfield  {journal} {\bibinfo  {journal} {Phys. Rev. B}\ }\textbf {\bibinfo
  {volume} {66}},\ \bibinfo {pages} {165104} (\bibinfo {year}
  {2002})}\BibitemShut {NoStop}%
\bibitem [{\citenamefont {Bisogni}\ \emph
  {et~al.}(2012{\natexlab{a}})\citenamefont {Bisogni}, \citenamefont
  {Simonelli}, \citenamefont {Ament}, \citenamefont {Forte}, \citenamefont
  {Moretti~Sala}, \citenamefont {Minola}, \citenamefont {Huotari},
  \citenamefont {van~den Brink}, \citenamefont {Ghiringhelli}, \citenamefont
  {Brookes},\ and\ \citenamefont {Braicovich}}]{Bisogni2012a}%
  \BibitemOpen
  \bibfield  {author} {\bibinfo {author} {\bibfnamefont {V.}~\bibnamefont
  {Bisogni}}, \bibinfo {author} {\bibfnamefont {L.}~\bibnamefont {Simonelli}},
  \bibinfo {author} {\bibfnamefont {L.~J.~P.}\ \bibnamefont {Ament}}, \bibinfo
  {author} {\bibfnamefont {F.}~\bibnamefont {Forte}}, \bibinfo {author}
  {\bibfnamefont {M.}~\bibnamefont {Moretti~Sala}}, \bibinfo {author}
  {\bibfnamefont {M.}~\bibnamefont {Minola}}, \bibinfo {author} {\bibfnamefont
  {S.}~\bibnamefont {Huotari}}, \bibinfo {author} {\bibfnamefont
  {J.}~\bibnamefont {van~den Brink}}, \bibinfo {author} {\bibfnamefont
  {G.}~\bibnamefont {Ghiringhelli}}, \bibinfo {author} {\bibfnamefont {N.~B.}\
  \bibnamefont {Brookes}}, \ and\ \bibinfo {author} {\bibfnamefont
  {L.}~\bibnamefont {Braicovich}},\ }\href {\doibase
  10.1103/PhysRevB.85.214527} {\bibfield  {journal} {\bibinfo  {journal} {Phys.
  Rev. B}\ }\textbf {\bibinfo {volume} {85}},\ \bibinfo {pages} {214527}
  (\bibinfo {year} {2012}{\natexlab{a}})}\BibitemShut {NoStop}%
\bibitem [{\citenamefont {Bisogni}\ \emph
  {et~al.}(2012{\natexlab{b}})\citenamefont {Bisogni}, \citenamefont
  {Moretti~Sala}, \citenamefont {Bendounan}, \citenamefont {Brookes},
  \citenamefont {Ghiringhelli},\ and\ \citenamefont
  {Braicovich}}]{Bisogni2012b}%
  \BibitemOpen
  \bibfield  {author} {\bibinfo {author} {\bibfnamefont {V.}~\bibnamefont
  {Bisogni}}, \bibinfo {author} {\bibfnamefont {M.}~\bibnamefont
  {Moretti~Sala}}, \bibinfo {author} {\bibfnamefont {A.}~\bibnamefont
  {Bendounan}}, \bibinfo {author} {\bibfnamefont {N.~B.}\ \bibnamefont
  {Brookes}}, \bibinfo {author} {\bibfnamefont {G.}~\bibnamefont
  {Ghiringhelli}}, \ and\ \bibinfo {author} {\bibfnamefont {L.}~\bibnamefont
  {Braicovich}},\ }\href {\doibase 10.1103/PhysRevB.85.214528} {\bibfield
  {journal} {\bibinfo  {journal} {Phys. Rev. B}\ }\textbf {\bibinfo {volume}
  {85}},\ \bibinfo {pages} {214528} (\bibinfo {year}
  {2012}{\natexlab{b}})}\BibitemShut {NoStop}%
\bibitem [{\citenamefont {Forte}\ \emph {et~al.}(2008)\citenamefont {Forte},
  \citenamefont {Ament},\ and\ \citenamefont {van~den Brink}}]{Forte2008}%
  \BibitemOpen
  \bibfield  {author} {\bibinfo {author} {\bibfnamefont {F.}~\bibnamefont
  {Forte}}, \bibinfo {author} {\bibfnamefont {L.~J.~P.}\ \bibnamefont {Ament}},
  \ and\ \bibinfo {author} {\bibfnamefont {J.}~\bibnamefont {van~den Brink}},\
  }\href {\doibase 10.1103/PhysRevB.77.134428} {\bibfield  {journal} {\bibinfo
  {journal} {Phys. Rev. B}\ }\textbf {\bibinfo {volume} {77}},\ \bibinfo {eid}
  {134428} (\bibinfo {year} {2008})}\BibitemShut {NoStop}%
\bibitem [{\citenamefont {Sugai}\ \emph {et~al.}(1988)\citenamefont {Sugai},
  \citenamefont {Shamoto},\ and\ \citenamefont {Sato}}]{Sugai1988}%
  \BibitemOpen
  \bibfield  {author} {\bibinfo {author} {\bibfnamefont {S.}~\bibnamefont
  {Sugai}}, \bibinfo {author} {\bibfnamefont {S.-i.}\ \bibnamefont {Shamoto}},
  \ and\ \bibinfo {author} {\bibfnamefont {M.}~\bibnamefont {Sato}},\ }\href
  {\doibase 10.1103/PhysRevB.38.6436} {\bibfield  {journal} {\bibinfo
  {journal} {Phys. Rev. B}\ }\textbf {\bibinfo {volume} {38}},\ \bibinfo
  {pages} {6436} (\bibinfo {year} {1988})}\BibitemShut {NoStop}%
\bibitem [{\citenamefont {R\"ubhausen}\ \emph {et~al.}(1997)\citenamefont
  {R\"ubhausen}, \citenamefont {Rieck}, \citenamefont {Dieckmann},
  \citenamefont {Subke}, \citenamefont {Bock},\ and\ \citenamefont
  {Merkt}}]{Ruebhausen1997}%
  \BibitemOpen
  \bibfield  {author} {\bibinfo {author} {\bibfnamefont {M.}~\bibnamefont
  {R\"ubhausen}}, \bibinfo {author} {\bibfnamefont {C.~T.}\ \bibnamefont
  {Rieck}}, \bibinfo {author} {\bibfnamefont {N.}~\bibnamefont {Dieckmann}},
  \bibinfo {author} {\bibfnamefont {K.-O.}\ \bibnamefont {Subke}}, \bibinfo
  {author} {\bibfnamefont {A.}~\bibnamefont {Bock}}, \ and\ \bibinfo {author}
  {\bibfnamefont {U.}~\bibnamefont {Merkt}},\ }\href {\doibase
  10.1103/PhysRevB.56.14797} {\bibfield  {journal} {\bibinfo  {journal} {Phys.
  Rev. B}\ }\textbf {\bibinfo {volume} {56}},\ \bibinfo {pages} {14797}
  (\bibinfo {year} {1997})}\BibitemShut {NoStop}%
\bibitem [{\citenamefont {Naeini}\ \emph {et~al.}(1999)\citenamefont {Naeini},
  \citenamefont {Chen}, \citenamefont {Irwin}, \citenamefont {Okuya},
  \citenamefont {Kimura},\ and\ \citenamefont {Kishio}}]{Naeini1999}%
  \BibitemOpen
  \bibfield  {author} {\bibinfo {author} {\bibfnamefont {J.~G.}\ \bibnamefont
  {Naeini}}, \bibinfo {author} {\bibfnamefont {X.~K.}\ \bibnamefont {Chen}},
  \bibinfo {author} {\bibfnamefont {J.~C.}\ \bibnamefont {Irwin}}, \bibinfo
  {author} {\bibfnamefont {M.}~\bibnamefont {Okuya}}, \bibinfo {author}
  {\bibfnamefont {T.}~\bibnamefont {Kimura}}, \ and\ \bibinfo {author}
  {\bibfnamefont {K.}~\bibnamefont {Kishio}},\ }\href {\doibase
  10.1103/PhysRevB.59.9642} {\bibfield  {journal} {\bibinfo  {journal} {Phys.
  Rev. B}\ }\textbf {\bibinfo {volume} {59}},\ \bibinfo {pages} {9642}
  (\bibinfo {year} {1999})}\BibitemShut {NoStop}%
\bibitem [{\citenamefont {Li}\ \emph {et~al.}(2013)\citenamefont {Li},
  \citenamefont {Le~Tacon}, \citenamefont {Matiks}, \citenamefont {Boris},
  \citenamefont {Loew}, \citenamefont {Lin}, \citenamefont {Chen},
  \citenamefont {Chan}, \citenamefont {Dorow}, \citenamefont {Ji},
  \citenamefont {Bari\ifmmode \check{s}\else \v{s}\fi{}i\ifmmode~\acute{c}\else
  \'{c}\fi{}}, \citenamefont {Zhao}, \citenamefont {Greven},\ and\
  \citenamefont {Keimer}}]{Li2013}%
  \BibitemOpen
  \bibfield  {author} {\bibinfo {author} {\bibfnamefont {Y.}~\bibnamefont
  {Li}}, \bibinfo {author} {\bibfnamefont {M.}~\bibnamefont {Le~Tacon}},
  \bibinfo {author} {\bibfnamefont {Y.}~\bibnamefont {Matiks}}, \bibinfo
  {author} {\bibfnamefont {A.~V.}\ \bibnamefont {Boris}}, \bibinfo {author}
  {\bibfnamefont {T.}~\bibnamefont {Loew}}, \bibinfo {author} {\bibfnamefont
  {C.~T.}\ \bibnamefont {Lin}}, \bibinfo {author} {\bibfnamefont
  {L.}~\bibnamefont {Chen}}, \bibinfo {author} {\bibfnamefont {M.~K.}\
  \bibnamefont {Chan}}, \bibinfo {author} {\bibfnamefont {C.}~\bibnamefont
  {Dorow}}, \bibinfo {author} {\bibfnamefont {L.}~\bibnamefont {Ji}}, \bibinfo
  {author} {\bibfnamefont {N.}~\bibnamefont {Bari\ifmmode \check{s}\else
  \v{s}\fi{}i\ifmmode~\acute{c}\else \'{c}\fi{}}}, \bibinfo {author}
  {\bibfnamefont {X.}~\bibnamefont {Zhao}}, \bibinfo {author} {\bibfnamefont
  {M.}~\bibnamefont {Greven}}, \ and\ \bibinfo {author} {\bibfnamefont
  {B.}~\bibnamefont {Keimer}},\ }\href {\doibase
  10.1103/PhysRevLett.111.187001} {\bibfield  {journal} {\bibinfo  {journal}
  {Phys. Rev. Lett.}\ }\textbf {\bibinfo {volume} {111}},\ \bibinfo {pages}
  {187001} (\bibinfo {year} {2013})}\BibitemShut {NoStop}%
\bibitem [{Sup()}]{Supple}%
  \BibitemOpen
  \href@noop {} {}\bibinfo {note} {See Supplemental Material for the details
  about theoretical two-magnon Raman scattering, dynamical DMRG, and RIXS
  spectra of LBCO ($x$ = 0.125) below and above the transition temperature of
  charge order.}\BibitemShut {Stop}%
\bibitem [{Mae()}]{Maekawa}%
  \BibitemOpen
  \href@noop {} {}\bibinfo {note} {S. Maekawa, T. Tohyama, S. E. Barnes, S.
  Ishihara, W. Koshibae, and G.Khaliullin: {\it Physics of Transition Metal
  Oxides} (Springer, Berlin, 2004) Springer Series in Solid-State Sciences,
  Vol. 144, Chapter 2.}\BibitemShut {Stop}%
\bibitem [{\citenamefont {Dean}\ \emph
  {et~al.}(2013{\natexlab{a}})\citenamefont {Dean}, \citenamefont {Dellea},
  \citenamefont {Springell}, \citenamefont {Yakhou-Harris}, \citenamefont
  {Kummer}, \citenamefont {Brookes}, \citenamefont {Liu}, \citenamefont {Sun},
  \citenamefont {Strle}, \citenamefont {Schmitt}, \citenamefont {Braicovich},
  \citenamefont {Ghiringhelli}, \citenamefont {Bo\v{o}vi\'c},\ and\
  \citenamefont {Hill}}]{Dean2013}%
  \BibitemOpen
  \bibfield  {author} {\bibinfo {author} {\bibfnamefont {M.~P.~M.}\
  \bibnamefont {Dean}}, \bibinfo {author} {\bibfnamefont {G.}~\bibnamefont
  {Dellea}}, \bibinfo {author} {\bibfnamefont {R.~S.}\ \bibnamefont
  {Springell}}, \bibinfo {author} {\bibfnamefont {F.}~\bibnamefont
  {Yakhou-Harris}}, \bibinfo {author} {\bibfnamefont {K.}~\bibnamefont
  {Kummer}}, \bibinfo {author} {\bibfnamefont {N.~B.}\ \bibnamefont {Brookes}},
  \bibinfo {author} {\bibfnamefont {X.}~\bibnamefont {Liu}}, \bibinfo {author}
  {\bibfnamefont {Y.-J.}\ \bibnamefont {Sun}}, \bibinfo {author} {\bibfnamefont
  {J.}~\bibnamefont {Strle}}, \bibinfo {author} {\bibfnamefont
  {T.}~\bibnamefont {Schmitt}}, \bibinfo {author} {\bibfnamefont
  {L.}~\bibnamefont {Braicovich}}, \bibinfo {author} {\bibfnamefont
  {G.}~\bibnamefont {Ghiringhelli}}, \bibinfo {author} {\bibfnamefont
  {I.}~\bibnamefont {Bo\v{o}vi\'c}}, \ and\ \bibinfo {author} {\bibfnamefont
  {J.~P.}\ \bibnamefont {Hill}},\ }\href {http://dx.doi.org/10.1038/nmat3723}
  {\bibfield  {journal} {\bibinfo  {journal} {Nat. Mater.}\ }\textbf {\bibinfo
  {volume} {12}},\ \bibinfo {pages} {1019} (\bibinfo {year}
  {2013}{\natexlab{a}})}\BibitemShut {NoStop}%
\bibitem [{\citenamefont {Dean}\ \emph
  {et~al.}(2013{\natexlab{b}})\citenamefont {Dean}, \citenamefont {Dellea},
  \citenamefont {Minola}, \citenamefont {Wilkins}, \citenamefont {Konik},
  \citenamefont {Gu}, \citenamefont {Le~Tacon}, \citenamefont {Brookes},
  \citenamefont {Yakhou-Harris}, \citenamefont {Kummer}, \citenamefont {Hill},
  \citenamefont {Braicovich},\ and\ \citenamefont {Ghiringhelli}}]{Dean2013c}%
  \BibitemOpen
  \bibfield  {author} {\bibinfo {author} {\bibfnamefont {M.~P.~M.}\
  \bibnamefont {Dean}}, \bibinfo {author} {\bibfnamefont {G.}~\bibnamefont
  {Dellea}}, \bibinfo {author} {\bibfnamefont {M.}~\bibnamefont {Minola}},
  \bibinfo {author} {\bibfnamefont {S.~B.}\ \bibnamefont {Wilkins}}, \bibinfo
  {author} {\bibfnamefont {R.~M.}\ \bibnamefont {Konik}}, \bibinfo {author}
  {\bibfnamefont {G.~D.}\ \bibnamefont {Gu}}, \bibinfo {author} {\bibfnamefont
  {M.}~\bibnamefont {Le~Tacon}}, \bibinfo {author} {\bibfnamefont {N.~B.}\
  \bibnamefont {Brookes}}, \bibinfo {author} {\bibfnamefont {F.}~\bibnamefont
  {Yakhou-Harris}}, \bibinfo {author} {\bibfnamefont {K.}~\bibnamefont
  {Kummer}}, \bibinfo {author} {\bibfnamefont {J.~P.}\ \bibnamefont {Hill}},
  \bibinfo {author} {\bibfnamefont {L.}~\bibnamefont {Braicovich}}, \ and\
  \bibinfo {author} {\bibfnamefont {G.}~\bibnamefont {Ghiringhelli}},\ }\href
  {\doibase 10.1103/PhysRevB.88.020403} {\bibfield  {journal} {\bibinfo
  {journal} {Phys. Rev. B}\ }\textbf {\bibinfo {volume} {88}},\ \bibinfo
  {pages} {020403} (\bibinfo {year} {2013}{\natexlab{b}})}\BibitemShut
  {NoStop}%
\bibitem [{\citenamefont {Wakimoto}\ \emph {et~al.}(2015)\citenamefont
  {Wakimoto}, \citenamefont {Ishii}, \citenamefont {Kimura}, \citenamefont
  {Fujita}, \citenamefont {Dellea}, \citenamefont {Kummer}, \citenamefont
  {Braicovich}, \citenamefont {Ghiringhelli}, \citenamefont {Debeer-Schmitt},\
  and\ \citenamefont {Granroth}}]{Wakimoto2015}%
  \BibitemOpen
  \bibfield  {author} {\bibinfo {author} {\bibfnamefont {S.}~\bibnamefont
  {Wakimoto}}, \bibinfo {author} {\bibfnamefont {K.}~\bibnamefont {Ishii}},
  \bibinfo {author} {\bibfnamefont {H.}~\bibnamefont {Kimura}}, \bibinfo
  {author} {\bibfnamefont {M.}~\bibnamefont {Fujita}}, \bibinfo {author}
  {\bibfnamefont {G.}~\bibnamefont {Dellea}}, \bibinfo {author} {\bibfnamefont
  {K.}~\bibnamefont {Kummer}}, \bibinfo {author} {\bibfnamefont
  {L.}~\bibnamefont {Braicovich}}, \bibinfo {author} {\bibfnamefont
  {G.}~\bibnamefont {Ghiringhelli}}, \bibinfo {author} {\bibfnamefont {L.~M.}\
  \bibnamefont {Debeer-Schmitt}}, \ and\ \bibinfo {author} {\bibfnamefont
  {G.~E.}\ \bibnamefont {Granroth}},\ }\href {\doibase
  10.1103/PhysRevB.91.184513} {\bibfield  {journal} {\bibinfo  {journal} {Phys.
  Rev. B}\ }\textbf {\bibinfo {volume} {91}},\ \bibinfo {pages} {184513}
  (\bibinfo {year} {2015})}\BibitemShut {NoStop}%
\bibitem [{\citenamefont {Monney}\ \emph {et~al.}(2016)\citenamefont {Monney},
  \citenamefont {Schmitt}, \citenamefont {Matt}, \citenamefont {Mesot},
  \citenamefont {Strocov}, \citenamefont {Lipscombe}, \citenamefont {Hayden},\
  and\ \citenamefont {Chang}}]{Monney2016}%
  \BibitemOpen
  \bibfield  {author} {\bibinfo {author} {\bibfnamefont {C.}~\bibnamefont
  {Monney}}, \bibinfo {author} {\bibfnamefont {T.}~\bibnamefont {Schmitt}},
  \bibinfo {author} {\bibfnamefont {C.~E.}\ \bibnamefont {Matt}}, \bibinfo
  {author} {\bibfnamefont {J.}~\bibnamefont {Mesot}}, \bibinfo {author}
  {\bibfnamefont {V.~N.}\ \bibnamefont {Strocov}}, \bibinfo {author}
  {\bibfnamefont {O.~J.}\ \bibnamefont {Lipscombe}}, \bibinfo {author}
  {\bibfnamefont {S.~M.}\ \bibnamefont {Hayden}}, \ and\ \bibinfo {author}
  {\bibfnamefont {J.}~\bibnamefont {Chang}},\ }\href {\doibase
  10.1103/PhysRevB.93.075103} {\bibfield  {journal} {\bibinfo  {journal} {Phys.
  Rev. B}\ }\textbf {\bibinfo {volume} {93}},\ \bibinfo {pages} {075103}
  (\bibinfo {year} {2016})}\BibitemShut {NoStop}%
\bibitem [{\citenamefont {Benjamin}\ \emph {et~al.}(2014)\citenamefont
  {Benjamin}, \citenamefont {Klich},\ and\ \citenamefont
  {Demler}}]{Benjamin2014}%
  \BibitemOpen
  \bibfield  {author} {\bibinfo {author} {\bibfnamefont {D.}~\bibnamefont
  {Benjamin}}, \bibinfo {author} {\bibfnamefont {I.}~\bibnamefont {Klich}}, \
  and\ \bibinfo {author} {\bibfnamefont {E.}~\bibnamefont {Demler}},\ }\href
  {\doibase 10.1103/PhysRevLett.112.247002} {\bibfield  {journal} {\bibinfo
  {journal} {Phys. Rev. Lett.}\ }\textbf {\bibinfo {volume} {112}},\ \bibinfo
  {pages} {247002} (\bibinfo {year} {2014})}\BibitemShut {NoStop}%
\bibitem [{\citenamefont {Tranquada}\ \emph {et~al.}(1995)\citenamefont
  {Tranquada}, \citenamefont {Sternlieb}, \citenamefont {Axe}, \citenamefont
  {Nakamura},\ and\ \citenamefont {Uchida}}]{Tranquada1995}%
  \BibitemOpen
  \bibfield  {author} {\bibinfo {author} {\bibfnamefont {J.~M.}\ \bibnamefont
  {Tranquada}}, \bibinfo {author} {\bibfnamefont {B.~J.}\ \bibnamefont
  {Sternlieb}}, \bibinfo {author} {\bibfnamefont {J.~D.}\ \bibnamefont {Axe}},
  \bibinfo {author} {\bibfnamefont {Y.}~\bibnamefont {Nakamura}}, \ and\
  \bibinfo {author} {\bibfnamefont {S.}~\bibnamefont {Uchida}},\ }\href
  {http://dx.doi.org/10.1038/375561a0} {\bibfield  {journal} {\bibinfo
  {journal} {Nature}\ }\textbf {\bibinfo {volume} {375}},\ \bibinfo {pages}
  {561} (\bibinfo {year} {1995})}\BibitemShut {NoStop}%
\bibitem [{\citenamefont {Ghiringhelli}\ \emph {et~al.}(2012)\citenamefont
  {Ghiringhelli}, \citenamefont {Le~Tacon}, \citenamefont {Minola},
  \citenamefont {Blanco-Canosa}, \citenamefont {Mazzoli}, \citenamefont
  {Brookes}, \citenamefont {De~Luca}, \citenamefont {Frano}, \citenamefont
  {Hawthorn}, \citenamefont {He}, \citenamefont {Loew}, \citenamefont {Sala},
  \citenamefont {Peets}, \citenamefont {Salluzzo}, \citenamefont {Schierle},
  \citenamefont {Sutarto}, \citenamefont {Sawatzky}, \citenamefont {Weschke},
  \citenamefont {Keimer},\ and\ \citenamefont {Braicovich}}]{Ghiringhelli2012}%
  \BibitemOpen
  \bibfield  {author} {\bibinfo {author} {\bibfnamefont {G.}~\bibnamefont
  {Ghiringhelli}}, \bibinfo {author} {\bibfnamefont {M.}~\bibnamefont
  {Le~Tacon}}, \bibinfo {author} {\bibfnamefont {M.}~\bibnamefont {Minola}},
  \bibinfo {author} {\bibfnamefont {S.}~\bibnamefont {Blanco-Canosa}}, \bibinfo
  {author} {\bibfnamefont {C.}~\bibnamefont {Mazzoli}}, \bibinfo {author}
  {\bibfnamefont {N.~B.}\ \bibnamefont {Brookes}}, \bibinfo {author}
  {\bibfnamefont {G.~M.}\ \bibnamefont {De~Luca}}, \bibinfo {author}
  {\bibfnamefont {A.}~\bibnamefont {Frano}}, \bibinfo {author} {\bibfnamefont
  {D.~G.}\ \bibnamefont {Hawthorn}}, \bibinfo {author} {\bibfnamefont
  {F.}~\bibnamefont {He}}, \bibinfo {author} {\bibfnamefont {T.}~\bibnamefont
  {Loew}}, \bibinfo {author} {\bibfnamefont {M.~M.}\ \bibnamefont {Sala}},
  \bibinfo {author} {\bibfnamefont {D.~C.}\ \bibnamefont {Peets}}, \bibinfo
  {author} {\bibfnamefont {M.}~\bibnamefont {Salluzzo}}, \bibinfo {author}
  {\bibfnamefont {E.}~\bibnamefont {Schierle}}, \bibinfo {author}
  {\bibfnamefont {R.}~\bibnamefont {Sutarto}}, \bibinfo {author} {\bibfnamefont
  {G.~A.}\ \bibnamefont {Sawatzky}}, \bibinfo {author} {\bibfnamefont
  {E.}~\bibnamefont {Weschke}}, \bibinfo {author} {\bibfnamefont
  {B.}~\bibnamefont {Keimer}}, \ and\ \bibinfo {author} {\bibfnamefont
  {L.}~\bibnamefont {Braicovich}},\ }\href {\doibase 10.1126/science.1223532}
  {\bibfield  {journal} {\bibinfo  {journal} {Science}\ }\textbf {\bibinfo
  {volume} {337}},\ \bibinfo {pages} {821} (\bibinfo {year}
  {2012})}\BibitemShut {NoStop}%
\bibitem [{\citenamefont {Comin}\ and\ \citenamefont
  {Damascelli}(2016)}]{Comin2016}%
  \BibitemOpen
  \bibfield  {author} {\bibinfo {author} {\bibfnamefont {R.}~\bibnamefont
  {Comin}}\ and\ \bibinfo {author} {\bibfnamefont {A.}~\bibnamefont
  {Damascelli}},\ }\href {\doibase 10.1146/annurev-conmatphys-031115-011401}
  {\bibfield  {journal} {\bibinfo  {journal} {Annu. Rev. Condens. Matter
  Phys.}\ }\textbf {\bibinfo {volume} {7}},\ \bibinfo {pages} {369} (\bibinfo
  {year} {2016})}\BibitemShut {NoStop}%
\bibitem [{\citenamefont {Tokura}\ \emph {et~al.}(1990)\citenamefont {Tokura},
  \citenamefont {Koshihara}, \citenamefont {Arima}, \citenamefont {Takagi},
  \citenamefont {Ishibashi}, \citenamefont {Ido},\ and\ \citenamefont
  {Uchida}}]{Tokura1990}%
  \BibitemOpen
  \bibfield  {author} {\bibinfo {author} {\bibfnamefont {Y.}~\bibnamefont
  {Tokura}}, \bibinfo {author} {\bibfnamefont {S.}~\bibnamefont {Koshihara}},
  \bibinfo {author} {\bibfnamefont {T.}~\bibnamefont {Arima}}, \bibinfo
  {author} {\bibfnamefont {H.}~\bibnamefont {Takagi}}, \bibinfo {author}
  {\bibfnamefont {S.}~\bibnamefont {Ishibashi}}, \bibinfo {author}
  {\bibfnamefont {T.}~\bibnamefont {Ido}}, \ and\ \bibinfo {author}
  {\bibfnamefont {S.}~\bibnamefont {Uchida}},\ }\href {\doibase
  10.1103/PhysRevB.41.11657} {\bibfield  {journal} {\bibinfo  {journal} {Phys.
  Rev. B}\ }\textbf {\bibinfo {volume} {41}},\ \bibinfo {pages} {11657}
  (\bibinfo {year} {1990})}\BibitemShut {NoStop}%
\end{thebibliography}%

\end{document}